# Understanding heat dissipation factors for fixed-tilt and single-axis tracked open-rack photovoltaic modules: experimental insights

**Johannes Pretorius**[1*], **Shaun Nielsen**[1]

[1]Department of Mechanical and Mechatronic Engineering, Stellenbosch University

Private Bag X1, Matieland, 7602, Stellenbosch, South Africa

[*]Corresponding author, jpp@sun.ac.za

**Abstract**

This paper presents the results of long-term experiments conducted on fixed-tilt (FT) and single-axis tracked (SAT) open-rack photovoltaic (PV) modules in South Africa. Utilizing Faiman's heat dissipation model and data filtering method, the study demonstrates favourable comparisons of FT experimental results with literature, while yielding novel heat dissipation factors for SAT modules. Enhanced heat dissipation is observed in no/low wind conditions for SAT modules compared to FT modules. Analyses reveal the influence of plane-of-array (POA) irradiance, wind speed and direction on module temperature, with SAT modules exhibiting greater heat dissipation stability. An investigation into data filtering methods suggests minor sensitivity for both configurations, with a slightly more pronounced impact on SAT modules. Assessments comparing module temperature predictions using diverse heat dissipation factors for FT modules reveal negligible sensitivity. This suggests that exact heat dissipation factor values may not be crucial for accurate predictions of module temperature in FT open-rack systems. Annual power output simulations using PVsyst software demonstrate a 2.9% and 3.3% enhancement for FT and SAT configurations, respectively, when employing experimentally determined heat dissipation factors. These findings highlight the importance of realistic, configuration-specific heat dissipation factors in optimizing PV system performance, particularly in the competitive context of modern PV power plant construction and techno-economic calculations.

# 1. Introduction

The power output of solar PV arrays is affected by their operational efficiency, which in turn depend on the temperatures experienced by the module cells. Heat dissipation from PV modules plays a major role in determining module temperature, with the effectiveness of heat dissipation in PV arrays being influenced by the installation configuration [1]. The accurate calculation of heat dissipation therefore merits careful consideration when designing and analysing PV systems.

Module efficiency reductions with temperature are universal, regardless of mounting configuration or module type. For each degree rise in module operating temperature, typical conversion efficiency reductions of 0.3% - 0.5% have been reported [2].

Various steady-state models for predicting PV module temperature have been proposed. Among these, an early study by Ross [3] determined a coefficient for relating module temperature to ambient temperature and POA irradiance. A study by Servant [4] developed a heat exchange model based on experimental measurements for module temperature, ambient temperature, POA irradiance and wind speed. From the site measurements, three coefficients are obtained to define his model. King et al. [5] measured the same parameters during their experiments. Testing considered open-rack and insulated modules and the model included coefficients to account for differences in configuration. An energy balance approach is followed by Mattei et al. [6] to produce an experimentally validated model which predicts module temperature and proposes relations for evaluation of an overall heat dissipation factor. Faiman [7] conducted experiments on seven different types of PV modules in open-rack configuration and proposed a simple model based on the Hottel-Whillier-Bliss equation [8, 9]. His method uses linear regression to obtain two heat dissipation constants that describe the heat transfer from the module to the environment. Semi-empirical explicit correlations are presented by Skoplaki et al. [10], together with a dimensionless mounting parameter to account for different setups. Work by Kaplani and Kaplanis [11] develop a thermal model and use experimental data to assess the dependence of module temperature on wind direction and speed, as well as module inclination and orientation. Santos et al. [12] presents a review paper on the available module temperature prediction models in literature. Kaplanis et al. [13] evaluates several of the semi-empirical models which have been proposed for the prediction of PV module temperature. They develop their own compact model which accounts for deviation of operating conditions from Standard Operating Conditions, efficiency, ageing, geometry, and cell technology. The model is employed to predict module temperatures for a range of configurations, including fixed-tilt and double-axis tracked open-rack systems.

The model by Faiman [7] has been widely adopted as basis for the calculation of module temperatures during PV system design and sizing. The commercial software PVSyst [14], for example, utilizes the Faiman model in its determination of heat dissipation from a PV array. The user is required to input specific heat dissipation factors based on the module configuration, whereafter the software calculates annual power output based on the annualized meteorological data at the location (solar irradiation, ambient temperature, and wind speed).

The PV module temperature according to this model is calculated from:

$$T_{mod} = T_{amb} + \left(\frac{H}{U_0' + U_1' \cdot V_w}\right) \qquad (1)$$

where the modified heat dissipation factors $U_0'$ and $U_1'$ are determined using:

$$U_0' = \frac{U_0}{\eta_0 - \eta_e} \qquad (2)$$

$$U_1' = \frac{U_1}{\eta_0 - \eta_e} \qquad (3)$$

In this context, $U_0$ is the constant heat dissipation factor, which encompasses the influence of radiation and natural convection heat transfer with the environment. The factor $U_1$, on the other hand, represents the wind-dependent heat dissipation factor. Lastly, $\eta_o$ and $\eta_e$ denote the optical and electrical efficiency of the PV module, respectively, while $H$ is the POA irradiance.

A number of research studies, utilizing Faiman's model, have performed experimental investigations to determine Faiman's heat dissipation factors for PV modules in various installation configurations. Most of this work has focussed on open-rack applications, while some building-integrated PV (BIPV) [13, 15, 17, 18], building-attached PV (BAPV) [13, 16, 17] and floating solar PV (FPV) [19] scenarios are also considered.

The focus of this paper is on PV module heat dissipation for FT and SAT open-rack applications, according to the Faiman approach. Considering this emphasis, the following contributions from literature for FT installations are noteworthy. Faiman [7] himself obtains values for $U_0'$ = 25 W/m$^2$K and $U_1'$ = 6.8 Ws/m$^3$K. A long-term experimental study by Koehl et al. [20] in the Negev desert and at an Alpine site in Germany attain average values of $U_0'$ = 27- 28 W/m$^2$K and $U_1'$ = 6-8 Ws/m$^3$K. The PV modules were tested using resistive loads that allow module operation near their maximum power point. Their study implemented a data filtering approach according to the IEC 61853-2 standard [21], which also adopts the Faiman method for the calculation of module temperature. In their investigation, Barykina and Hammer [22] assessed the heat dissipation factors at five distinct locations where open-rack modules were deployed. Their observations, conducted over a span of six months, indicated the following variations in heat dissipation factors for poly-crystalline silicon modules: $U'_0$ = 30- 42 W/m²K and $U'_1$ = 3-8 Ws/m³K.

There is a lack of available literature that addresses heat dissipation factors specifically for SAT open-rack PV configurations.

It is also worth noting that PVsyst employs default (unaccented) heat dissipation factors of $U_0$ = 29 W/m²K and $U_1$ = 0 Ws/m³K for any open-rack installation. The $U_0$ is effectively inflated to account for the contribution of wind, due to the potential unreliability of site-specific wind data.

As seen in the preceding information, even the literature on a single PV configuration (FT open-rack) presents a spectrum of values for heat dissipation factors. This seems to suggest the case-specific nature of these factors, or alternatively a sensitivity of the values to certain conditions, test details or even data processing methods.

This paper undertakes experiments at the same site to compare FT and SAT open-rack PV configurations. It introduces a novel aspect by developing heat dissipation factors specifically for SAT configurations. Additionally, the paper conducts an in-depth data analysis with the goal of quantifying the interplay between various variables that impact heat dissipation factors. Furthermore, it assesses the sensitivities of heat dissipation factors to daily fluctuations and evaluates the impact of different data filtering methods.

## 2. Experimental methodology

This section describes details of the test site, experimental setup, equipment, method, and data processing.

### 2.1 Test site

All tests were conducted at the Stellenbosch University Renewable Energy Center (SUNREC) facility, located approximately 10 km outside Stellenbosch, South Africa. Experimental measurements for several solar photovoltaic and solar thermal projects are hosted at the facility. The site consists of open farmland, which is ideal for measurements on PV arrays in open-rack configuration.

## 2.2 Experimental layout and equipment

Figure 1 shows the layout of the experiment at SUNREC, where the PV modules installed in SAT (modules 1 and 2) and FT (modules 3 and 4) configurations are depicted. Modules 1 and 2 were mounted on the available space of a single-axis tracker forming part of other ongoing measurements on site. This tracker runs North-South, meaning that tracking rotates from East to West during the day. The FT modules 3 and 4 were mounted on a structure with fixed tilt, facing due North. Table 1 offers a summary of the test site coordinates (including height above mean sea level (HAMSL) and height above ground level (HAGL)), PV module specifications, and setup details for the two mounting configurations.

During the testing period, PV module temperature, meteorological conditions, and SAT position data was monitored at the site. A schematic of the equipment position and connections is presented in Figure 2.

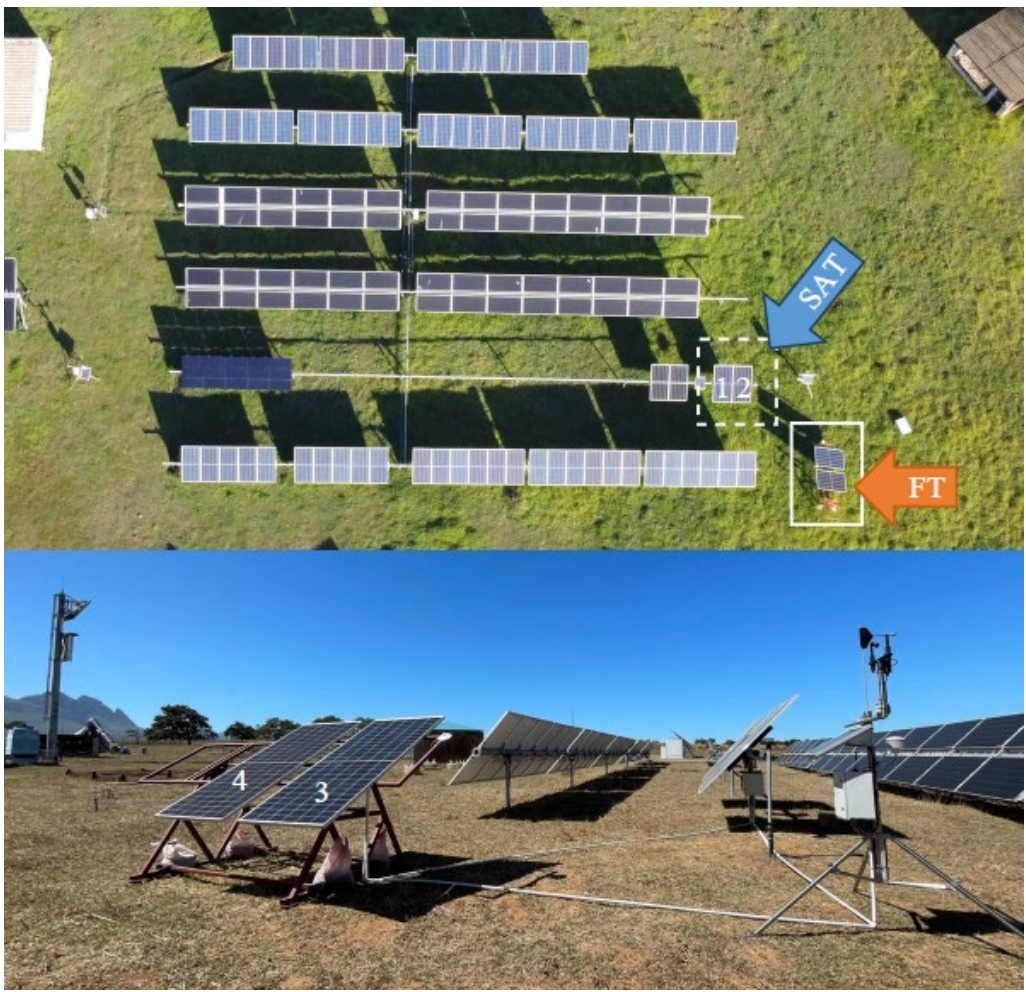

**Figure 1: Layout of experiment at SUNREC**

**Table 1: Test site, PV module specification and setup overview**

| Parameter | Specification / value | |
|---|---|---|
| PV module | CS3W-420P | |
| PV cell type | Poly-crystalline | |
| Module dimensions | 2108x1048x40 mm | |
| Nominal max. power | 420 W | |
| | | |
| Latitude | -33.853427° | |
| Longitude | 18.823940° | |
| HAMSL | 177 m | |
| HAGL | 1.5 m | |
| | | |
| Configuration | FT | SAT |
| No. of modules | 2 | 2 |
| Orientation | Portrait | Landscape |
| Tracking | - | East-West |
| Tilt angle | 31° | 0° |
| Module azimuth | 0° | 0° |

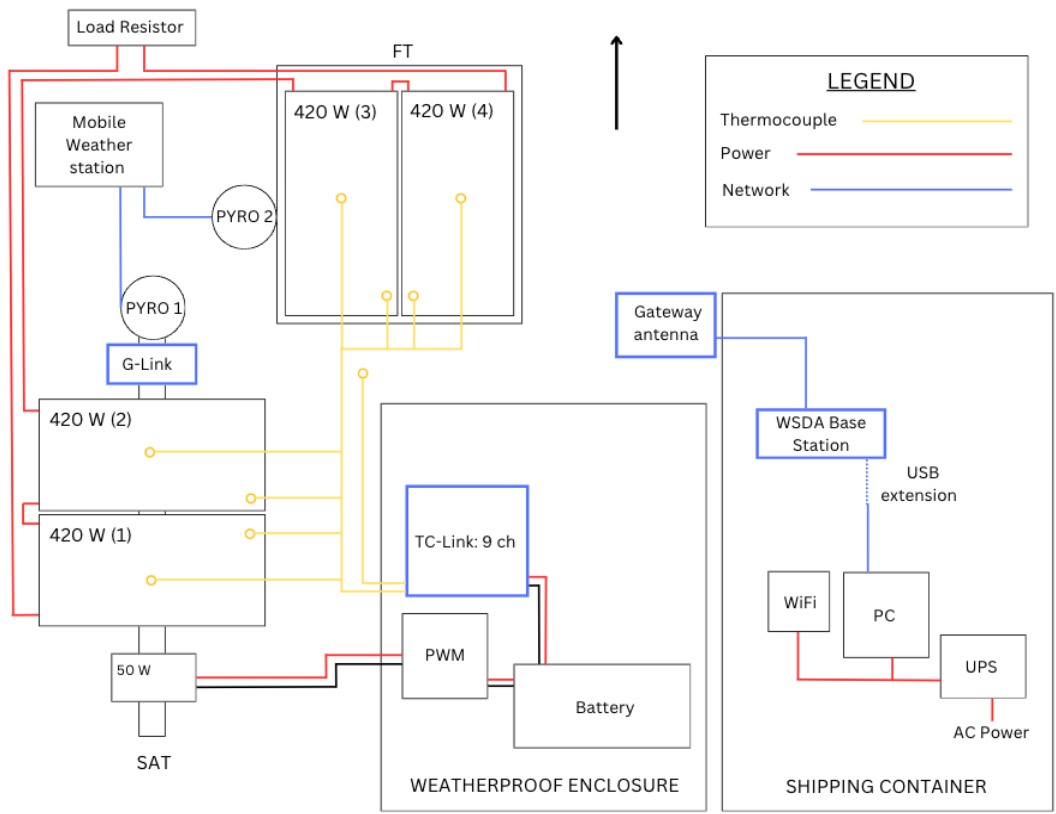

**Figure 2: Schematic of equipment position and connections**

The measurement system was selected to provide wireless transmission of data, remote monitoring and access to data, as well as data storage. PV module temperature was measured at the central and corner cells of each module, using two T-type thermocouples. The exposed junctions of the thermocouples were affixed to the back surface of each module using aluminium tape. Another shielded T-type thermocouple was used as a secondary measurement of ambient temperature. These nine thermocouples connected to a 12-channel Lord TC-link-200 temperature sensor node, which logs and transmits the data wirelessly to a Lord WSDA base station. The TC-link node operates using lithium batteries, with a 12 V backup battery. A 50 W solar panel charges the backup battery via a Pulse Width Modulator (PWM) charge controller. The node and its power supply were fitted in a weatherproof enclosure close to the PV modules.

A Lord G-link 200 accelerometer node was installed on the SAT torque tube. This sensor logged and transmitted the tilt angle of the SAT modules wirelessly to the base station. The tracker undergoes a continuous adjustment process, starting horizontally in the morning and making subsequent adjustments at five-minute intervals throughout the day. This sequential modification results in a linear alteration of the tracker's angle, allowing it to continually align with the sun's position. When shading is detected, the tracker readjusts to a relatively horizontal position during the nighttime hours.

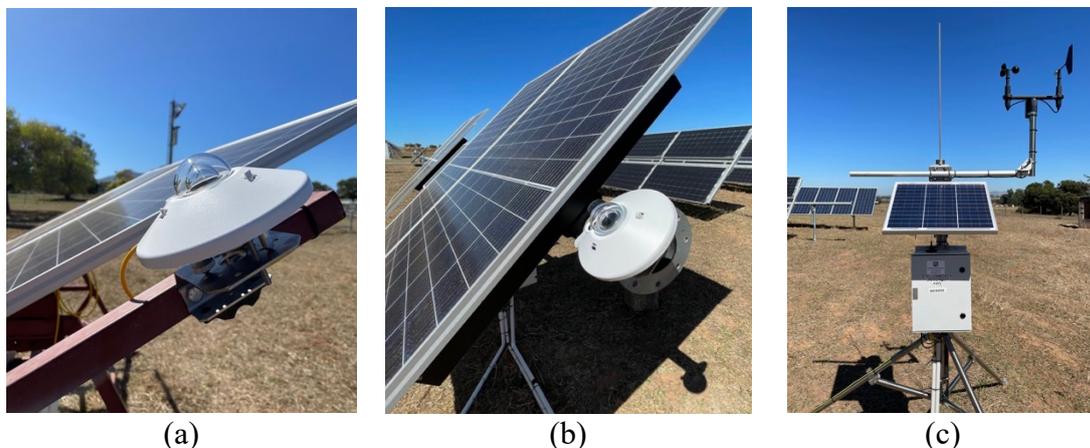

(a)   (b)   (c)
**Figure 3: (a) In-plane pyranometer for FT modules, (b) In-plane pyranometer for SAT modules, (c) Weather station**

Solar irradiance, ambient temperature, and wind speed were measured using a Campbell Scientific portable CR310 weather station [23]. Figure 1 and Figure 3 (c) show the weather station, which was installed approximately 5 m away from each mounting structure. The primary ambient temperature measurement was recorded using a shielded HygroVUE5 digital temperature sensor. Wind speed and direction measurements were obtained via a R.M. Young 03002 wind sentry and vane. The POA irradiance was determined by positioning Kipp & Zonen

CMP10 pyranometers in-plane with the FT and SAT structures, as depicted in Figure 3 (a) and (b) respectively. Further specifications of the instruments used to measure the experimental variables are listed Table 2.

Table 2: Experimental variables and corresponding instrument specifications

| Variable | Sensor | Accuracy | Range |
|---|---|---|---|
| $H$ | Kipp & Zonen CMP10 | 7 - 14 µV/W/m$^2$ | Measuring: 0 to 4000 W/m$^2$ Operating: -40 to 80 °C Spectral: 285 to 2800 nm |
| $V_w$ | R.M. Young 03002 | ±0.5 m/s | Measuring: 0 to 50 m/s Operating: -50 to 50 °C |
| $T_{mod}$ | T-type thermocouple | ±1 °C | -200 to 200 °C |
| $T_{amb}$ | HygroVUE5 | ±0.4 °C | -40 to 70 °C |

Maximum power point tracking (MPPT) charge controllers were not at our disposal for this investigation; however, load resistors were available. Although load resistors cannot guarantee that the PV modules function precisely at their maximum power point, they do ensure that the modules operate as close to real-world installations as our available experimental equipment allows. The connection of modules to load resistors also facilitates comparisons with previous research that utilized resistive loads in their experimental setups. We connected the load resistors in series to our four PV modules.

The base station facilitates communication between the node and a personal computer (PC). It manages the wireless data transmission within the network and enables real-time data visualization and up-to-date data collection. It connects to the PC via Universal Serial Bus (USB) and interfaces with compatible software, which stores the recorded data for remote access.

The PC was situated within a repurposed shipping container on site, positioned approximately 90 m away from the modules. To enable remote data collection and system monitoring, a mobile router was linked to the PC, which had remote access capabilities. An Uninterruptible Power Supply (UPS) was installed to guarantee continuous power supply and data collection.

The weather data was gathered using Campbell Scientific's CR300 datalogger. The PC400 datalogger support software was used for unit programming and remote data collection via PC and the LoggerLink app allowed for data collection directly to a mobile phone.

**2.3 Test method**

All instruments were calibrated prior to testing. After commencement of the experiment, data was collected 24 hours a day. Instruments recorded averaged values over 1-minute intervals. Data was accessed remotely, while periodic site visits were done to check the equipment and setup condition.

**2.4 Data processing**

The module temperature models developed for steady state analysis typically determine parameters used to model a representative value of the average temperature across a module and an average across an array in the case of power output simulations. In this work, the module temperature was calculated by averaging the two thermocouple measurements for each module. Fluctuations in conditions during testing can influence the accuracy of a model and filtering is applied to limit the analysis to times of relatively stable conditions. Such filtering is standard practice for thermal measurements on PV modules and the IEC 61853-2 standard [21] has defined measurement and filtering requirements.

The main concern of this study and associated results was developing heat dissipation factors according to the Faiman model, and as such the decision was made to implement the filtering strategy originally deployed by Faiman [7], with data points taken as averaged values over 5-minute intervals with the following filters:

- Clear sky days – no irradiance fluctuation during analysis period.
- Analysis between 10:00 and 14:00.

The reason for this period of data selection was to reduce noise in the data resulting from dividing small numbers by small numbers [7]. This occurrence would readily occur early or late in the day with smaller temperature differences. Before and after this period it is also expected to see larger changes to irradiance which is not suitable for a steady state model.

In addition to the above strategy from [7], the IEC standard [21] was used as a guideline, and the following criteria were also incorporated into the data processing procedure:

- Acceptable analysis should consist of data points from at least ten different days, each of those days should have at least ten data points before and after solar noon.

- Use 5-minute average wind speeds when plotting. A range of data points along the x-axis should cover at least 4 m/s.
- From all acceptable data points, calculate the average module temperature and plot $H/(T_{mod} - T_{amb})$ as a function of the 5-minute average wind speed. Use linear regression to determine the slope and intercept ($U'_1$ and $U'_0$ respectively) of the model.
- Report dates of experiment and the heat dissipation factors $U'_0$ and $U'_1$.

The standard [21] encompasses a broader set of criteria for data collection, data filtering, and evaluation details when compared to the methodology employed in this research. Our equipment and data sampling method does, however, provide sufficient accuracy to be comparable with testing according to the letter of the standard. The study's approach to filtering does not deviate from the idea of presenting steady state data. Rather, it is in keeping with Faiman's original work, with valuable criteria from the standard [21] adding quality to the data set while being practically achievable within the scope of this study.

The standard [21] goes on by noting that heat dissipation factors are ultimately used in the calculation of annualized energy yield and advises to check the coefficient impact on yields, and not only considering parameter uncertainty. This comparison will be conducted in Section 4, wherein we will assess the impact of the acquired heat dissipation factors on PV annual energy yield simulations through the utilization of PVsyst.

## 3. Experimental results and discussion

This section presents the experimental results. Initially, a normalized distribution of the main experimental variables is given to provide context on the range of environmental conditions experienced during testing. Thereafter, general trends of the measured parameters are shown. Subsequently, we present the heat dissipation factors acquired for both FT and SAT configurations, followed by an exploration of the distinctions and the corresponding heat transfer mechanisms. The sections that follow explore the relationships between irradiance, wind speed and module temperature, and investigate the sensitivity of heat dissipation factor values to the method of data filtering and daily variations.

### 3.1 Environmental conditions

Histograms of wind speed ($V_w$), ambient temperature ($T_{amb}$) and POA irradiance ($H$) are shown in Figure 4 and Figure 5 for the respective FT and SAT data sets.

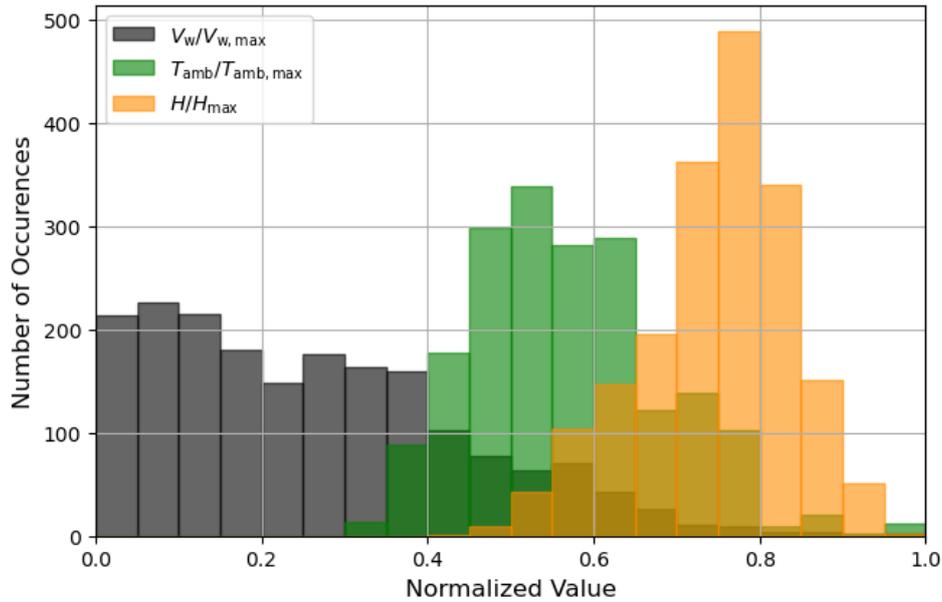

**Figure 4: Histogram of measured parameters for FT configuration**

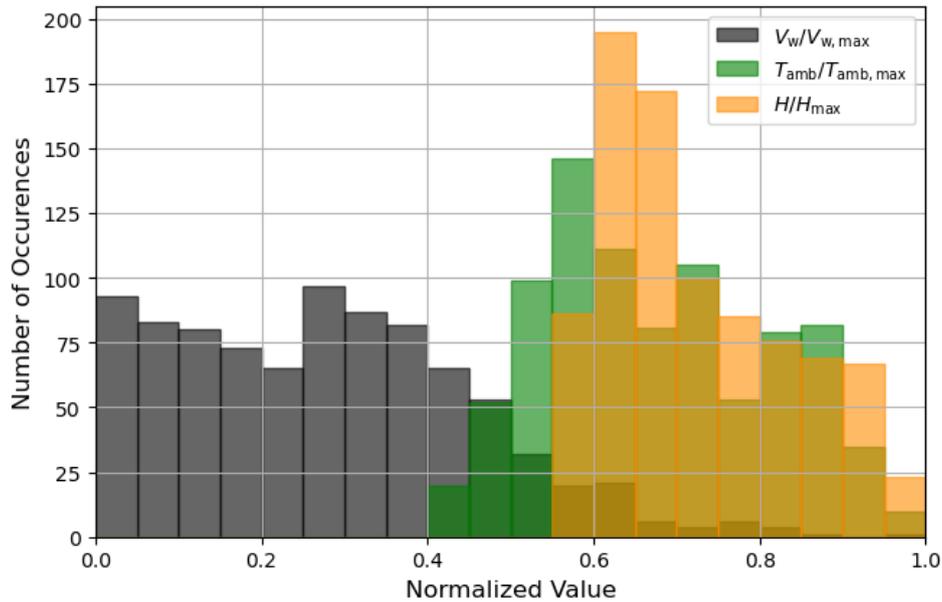

**Figure 5: Histogram of measured parameters for SAT configuration**

Data was collected during the period 30/03/2023 to 07/08/2023, where 44 days of a possible 131 met the criteria specified for a clear sky day. Due to tracker limitations, only 20 days proved satisfactory for inclusion into the final SAT data set. Normalizations are done using the following maximum parameter values: $V_{w,max} = 6.37 \ m/s$, $H_{max} = 1175.2 \ W/m^2$, $T_{amb,max} = 32.8°C$ for the FT data set and $V_{w,max} = 6.19\ m/s$, $H_{max} = 996.7\ W/m^2$, $T_{amb,max} = 28.4°C$ for the SAT data set.

From Figure 4 and Figure 5, the distributions for wind speed and ambient temperature are predictably similar for measurements at the same site, except for slight variations due to the difference in final data set between the configurations. The histogram for POA irradiance for the FT modules shows a relatively symmetrical distribution, compared to the left-skewed distribution of the SAT configuration. These distributions are functions of the typical daily variation of $H$ for the two different configurations (see Figure 6), the irradiance conditions during the seasons when testing occurred and the variation in comparative final data set.

### 3.2 Daily trends

Figure 6 presents a comparison of the POA irradiance experienced by the FT and SAT modules on a typical test day.

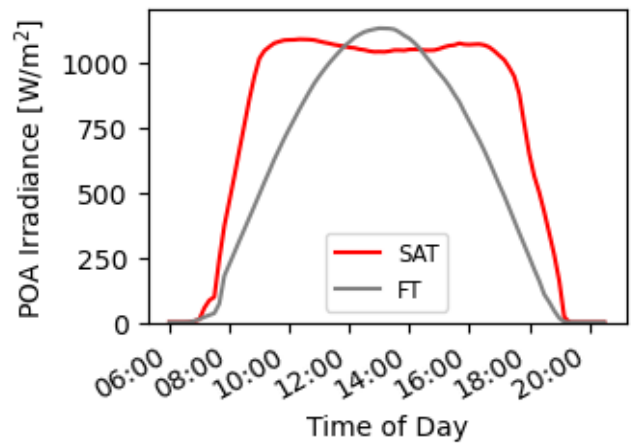

**Figure 6: Comparative plot of POA irradiance on a typical test day**

The impact of the tracker, which maintains high irradiance on the surfaces of the SAT modules throughout a significant portion of the day, is clearly evident. Despite the irradiance enhancement provided by the tracker, the FT modules still experience greater daily maximum POA irradiance due to their North-facing orientation.

Experimental measurements for three selected days of the testing period are shown in Figure 7 to Figure 9. Data is presented for the isolated test window between 10:00 and 14:00, as previously discussed.

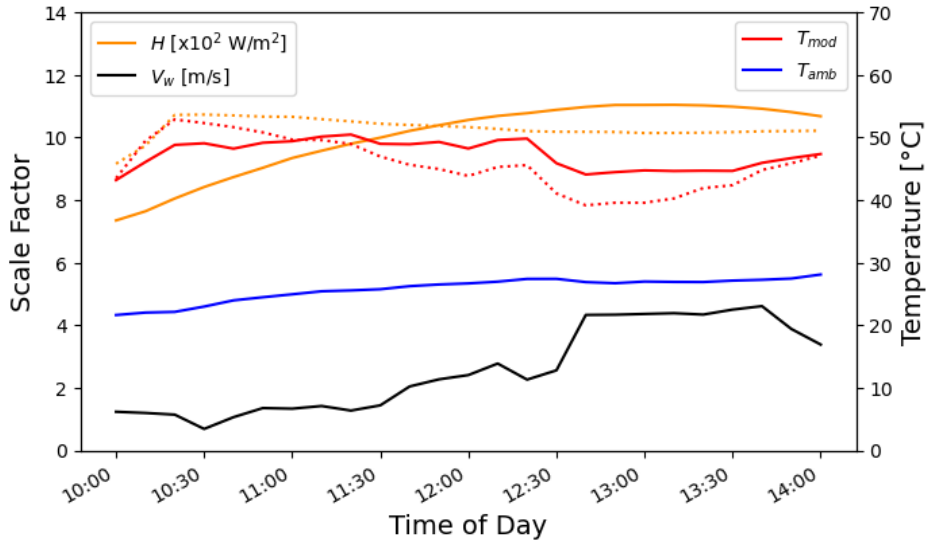

**Figure 7: Plot comparison of the experimental conditions on 13/03/2023 for the FT (solid lines) and SAT (dotted lines) configurations**

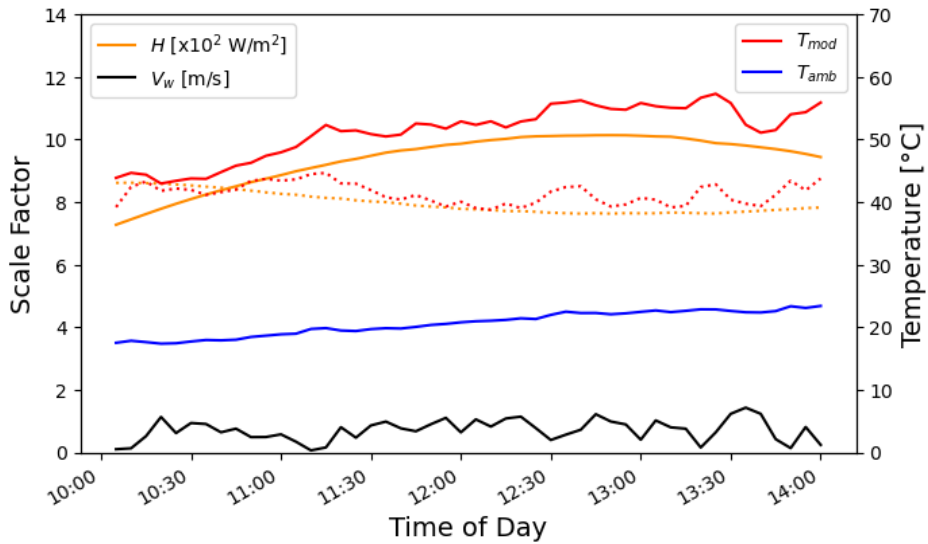

**Figure 8: Plot comparison of the experimental conditions on 30/04/2023 for the FT (solid lines) and SAT (dotted lines) configurations**

A number of trends are visible from these plots. In Figure 7, the module temperature demonstrates an inverse relationship with the wind speed profile, decreasing as the wind speed increases and vice versa as the wind speed reduces. The effect of POA irradiance on module temperature is also evident. Earlier in the day, the FT module temperature is lower than the SAT module temperature, while later in the day this trend is reversed. The relative module temperatures align with the magnitude of the POA irradiance experienced on the respective FT and SAT modules.

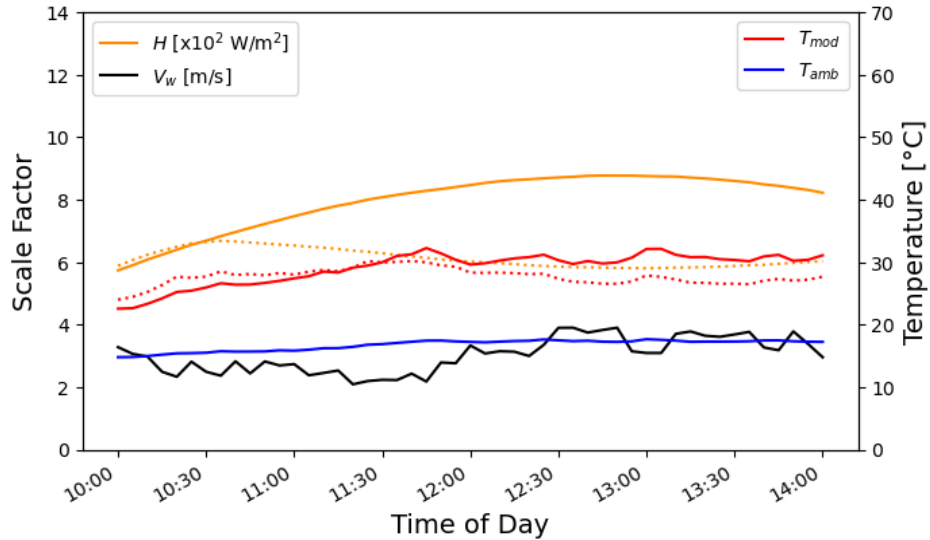

**Figure 9: Plot comparison of the experimental conditions on 25/06/2023 for the FT (solid lines) and SAT (dotted lines) configurations**

Figure 8 shows that, during a day of low wind speed, the module temperatures follow the trends of the corresponding POA irradiance curves closely. Under these conditions, module temperature differences of approximately 10-15°C between the FT and SAT modules are observed. In contrast, for a day with similar irradiance trends but higher wind speed, as in Figure 9, module temperatures differences between the configurations are minimal.

The daily trends therefore unmistakably affirm the anticipated significant impact of irradiance on module temperatures, as it governs the heat flux absorbed by the modules. Simultaneously, wind speed plays a crucial role in facilitating heat dissipation from the modules, contributing to the observed temperature variations.

### 3.3 Heat dissipation of FT and SAT configurations

Figure 10 summarizes the experimental data for the FT configuration. In this figure, Equation 1 is manipulated to plot $H/(T_{mod} - T_{amb})$ versus wind speed, which represents the effective heat dissipation characteristic of the modules. Over the testing period, 3378 5-minute averaged data points were collected between 10:00 and 14:00. An increasing trend is observed, which indicates the expected enhancement in heat dissipation from the modules with wind speed. Higher levels of scatter at higher wind speeds indicate that heat dissipation is more sensitive under these conditions, potentially also due to some wind speed fluctuations occurring at such times.

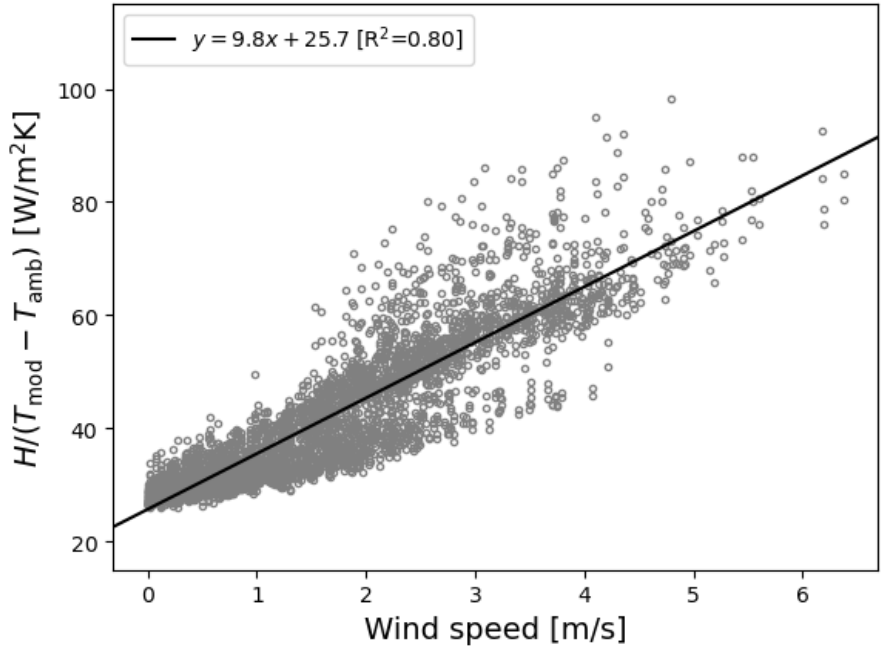

**Figure 10: Plot of H/(T$_{mod}$ − T$_{amb}$) versus wind speed for the FT configuration, and the associated linear fit for all clear sky 5-minute data points**

Performing linear regression using the method of least squares on the collected data enabled the determination of the $U_0'$ and $U_1'$ heat dissipation factors. The obtained values of $U_0'$ = 25.7 W/m²K and $U_1'$ = 9.8 Ws/m³K compare well to those initially determined by Faiman [7] ($U_0'$ = 25 W/m²K, $U_1'$ = 6.8 Ws/m³K, $R^2$ = 0.63). The $R^2$ value here is higher but has included data from more clear sky days over a longer period. Owen et al. [24] attained open-rack heat dissipation factors of $U_0'$ = 26.5 W/m²K and $U_1'$ = 5.2 Ws/m³K at a tilt angle of 23°. Their experimental site was located on a rooftop solar laboratory while the current SUNREC site is open farmland at ground level. This suggests that $U_1'$ is sensitive to the local wind conditions experienced at the test site.

A mean module temperature prediction was also produced from the above data, which resulted in a Root Mean Square Error (RMSE) of 2.55°C. Faiman [7] and Owen et al. [24] reported a RMSE of 1.86°K and 3.72°C respectively. Barykina and Hammer [22] conducted a validation using 15-minute averages which resulted in an average RMSE of 1.94°C for the polycrystalline modules across five test sites. The RMSE magnitude for the FT configuration of this study is therefore comparable to previous research.

Turning to the SAT configuration in Figure 11, 1746 acceptable data points were obtained for 5-minute averaged intervals between 10:00 and 14:00. The SAT results exhibit an improved correlation to a linear trend ($R^2 = 0.85$) compared to the FT setup, with consistent heat dissipation increases with higher wind speed. Notably, there is an absence of significant scatter at higher wind speeds in the SAT plot.

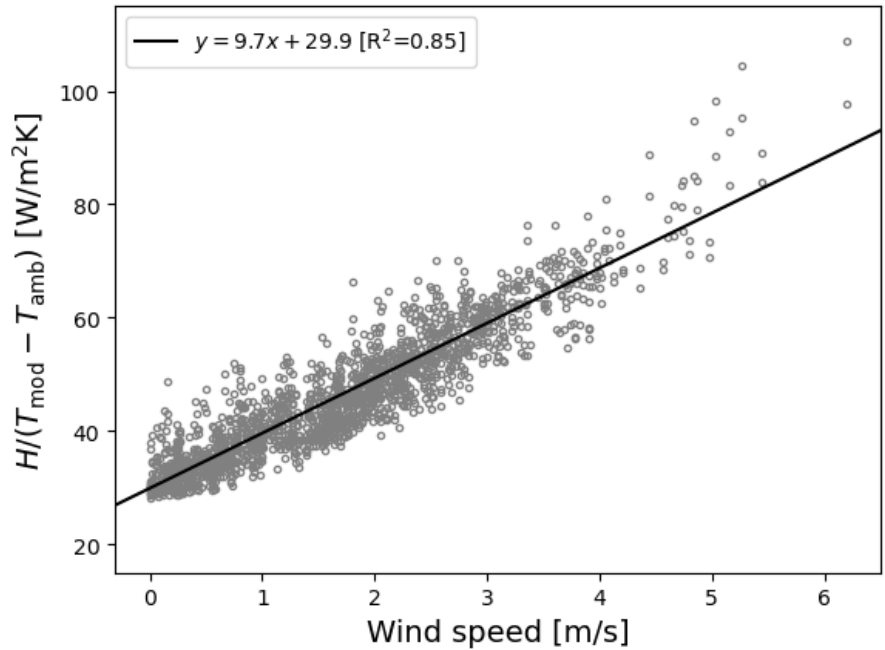

**Figure 11: Plot of H/(T$_{mod}$ − T$_{amb}$) versus wind speed for the SAT configuration, and the associated linear fit for all clear sky 5-minute data points**

Extracted from the SAT data are heat dissipation factors of $U'_0$ = 29.9 W/m²K and $U'_1$ = 9.7 Ws/m³K, which achieve an improved RMSE of 1.44°C compared to the FT scenario. The similarity in $U'_1$ values between the SAT and FT configurations implies that the determination of this value is significantly influenced by site conditions and the comparable level of exposure from the open-racked FT and SAT modules. The achieved $U'_0$ value is higher than for the FT configuration. Given the absence of prior literature on SAT configurations, further discussion will compare these results with those of the FT configuration.

As previously mentioned, the SAT configuration produced less usable data than the FT configuration. To enable a direct comparison, the FT data set was restricted to the same 1746 intervals utilized for the SAT evaluation. This adjustment did not significantly impact the values of the heat dissipation factors, with the values changing slightly to $U'_0$ = 25.6 W/m²K and $U'_1$ = 10.3 Ws/m³K for the FT modules.

The time average plots of Figure 12 to Figure 14 show the average value of the variable in the data set at each time over the duration of the limited comparable time period, with the maximum and minimum values included, for both the FT and SAT configurations. The trends of Figure 12 are the same as those presented in Figure 6, but limited to times between 10:00 and 14:00.

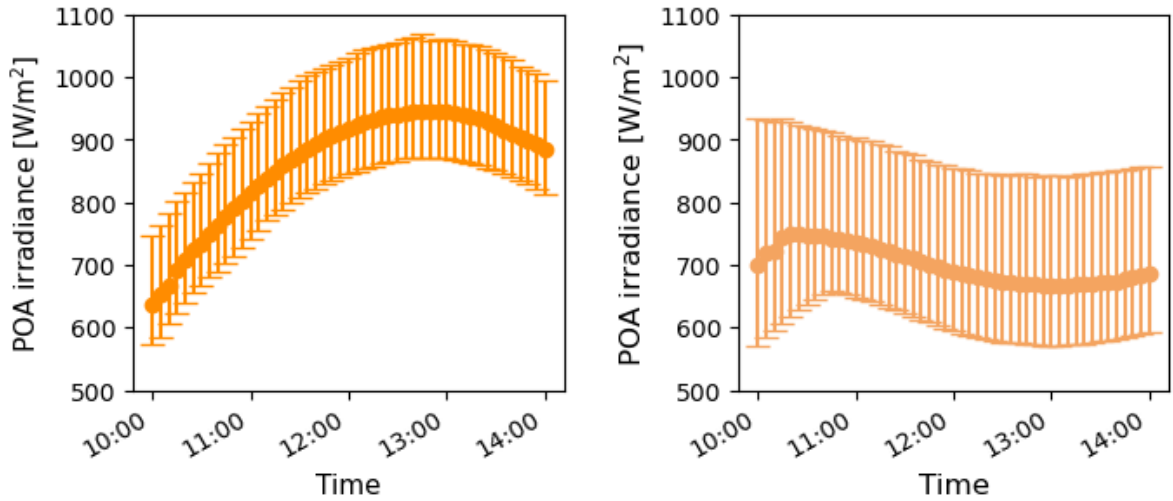

**Figure 12: Time average plot of POA irradiance for the FT (left) and SAT modules (right)**

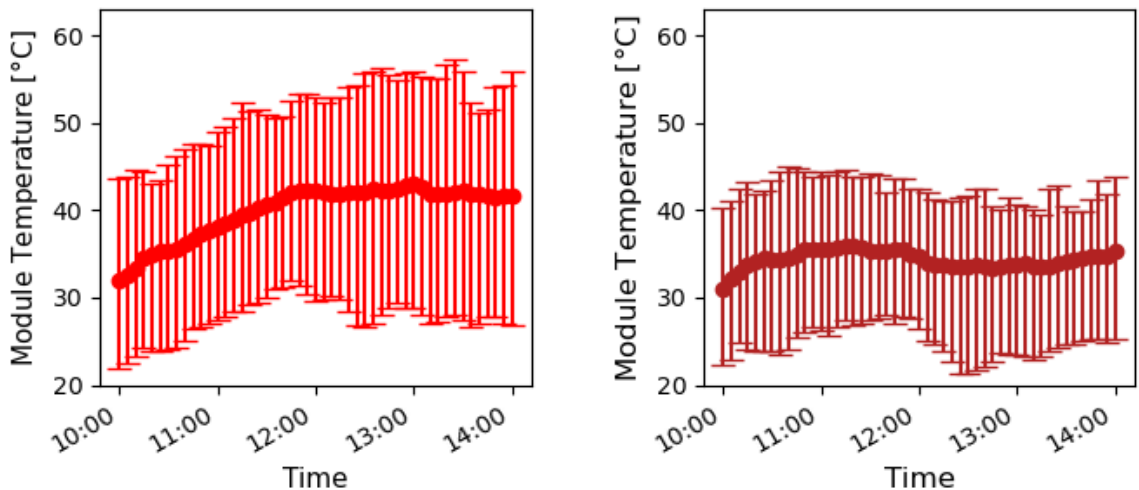

**Figure 13: Time average plot of module temperature for the FT (left) and SAT modules (right)**

The module temperature trends of Figure 13 plainly follow the corresponding irradiance trends of Figure 12, and show that the temperatures experienced by the FT modules are consistently higher than those of the SAT modules.

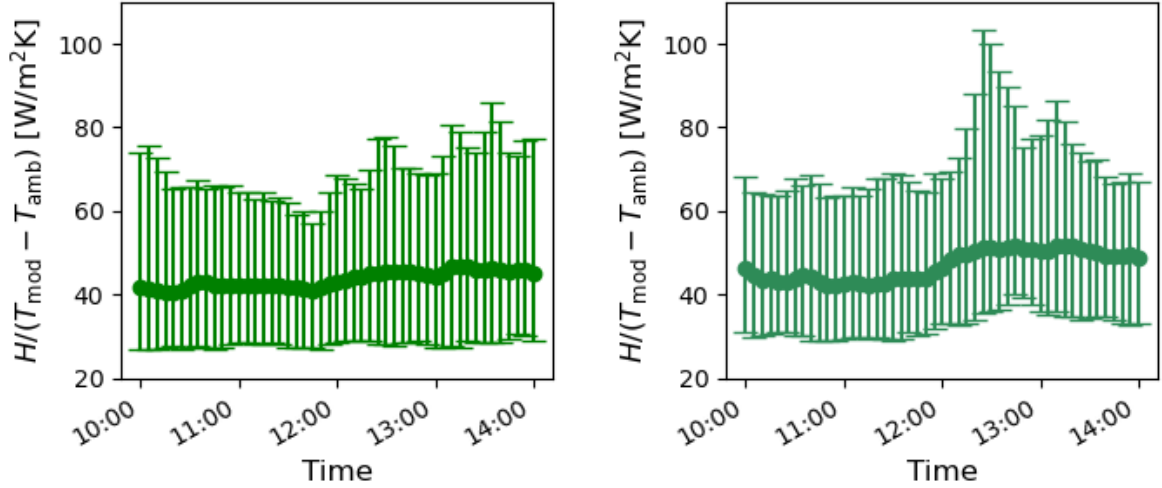

Figure 14: Time average plot of effective heat dissipation for the FT modules (left) and the SAT modules (right)

As noted in Section 3.2, the main drivers of module temperature are POA irradiance and wind speed. From Figure 12 it is clear that the POA irradiance on the East-West tracking SAT modules is very consistent, but significantly lower compared to the North-facing FT modules between 10:00 and 14:00 over the test period. The respective FT and SAT wind-dependant heat dissipation factors of $U'_1$ = 9.7 Ws/m³K and $U'_1$ = 10.3 Ws/m³K are comparable, indicating a similar forced convection cooling effect for both configurations during times of significant wind speed. Under no / low wind conditions, the higher $U'_0$ = 29.9 W/m²K factor of the SAT modules versus $U'_0$ = 25.7 W/m²K for the FT modules indicates enhanced heat dissipation for the tracking modules. The SAT modules spend significant time between 10:00 and 14:00 at lower inclination angles than the FT modules. In terms of natural convection, at lower inclination angles the front surface of the SAT modules would experience rising thermals, while the front surface of the FT modules would experience a stable boundary layer as the air rises over the inclined surface. This would produce a higher heat transfer rate for the SAT modules. The rear surfaces of both the SAT and FT modules would experience a stable boundary layer, while the heat dissipation from the SAT rear surface would be lower due to the low inclination. Radiation losses from the SAT rear surface to a cooler ground surface would be higher than for the FT modules, due to a higher view factor with the ground.

Figure 15 indicates that wind direction also plays a role in the heat dissipation characteristics of the FT and SAT modules. In this figure, a Northerly wind is denoted by 0° and a Southerly wind by 180°. The prevailing winds on site are generally from a Southerly or Westerly (270°) direction.

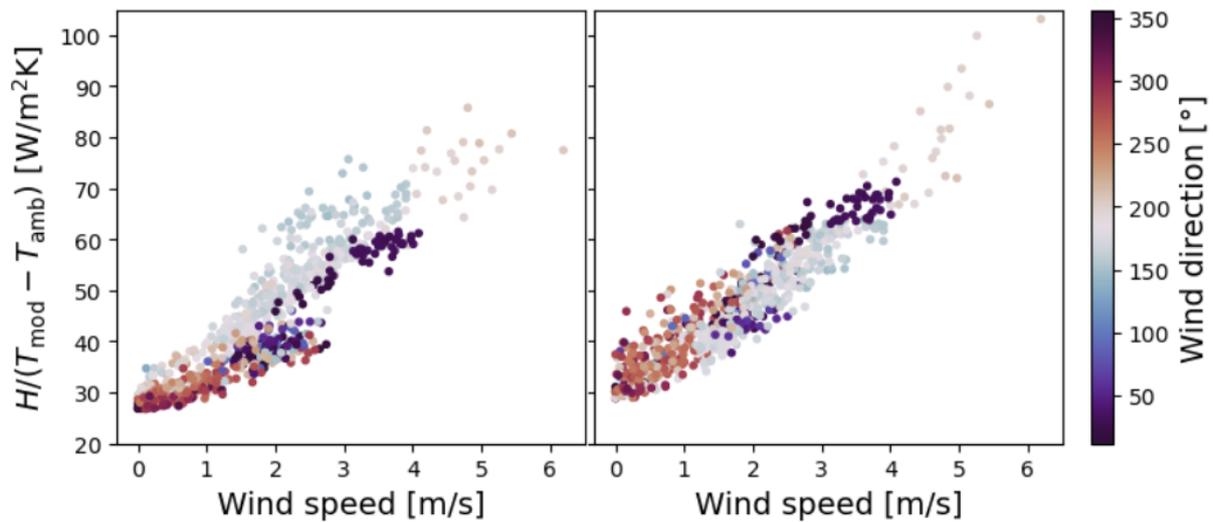

**Figure 15: Colour bar of wind direction over the plot of H/(T$_{mod}$ - T$_{amb}$) versus wind speed for the FT (left) and the SAT (right) configurations**

The left-hand side of Figure 15 indicates that greater heat dissipation is achieved from the FT modules with the wind from the South compared to the North. This finding correlates with the numerical and experimental observations of Glick et al. [25] for PV arrays. During times of higher winds from the South, the wind blows straight onto the back surface of the FT modules, while a Northerly wind blows into the front surface of the modules. Based on the insights by [25], it is expected that for Southerly winds, the flow will accelerate over the back face as it is forced underneath the module, while significant flow separation and recirculation will occur on the front face. During a North wind, little flow acceleration is expected over the front face, as the surface faces upward, with some recirculation and separation of flow over the back face. The acceleration and heightened mixing effects induced by Southerly winds consequently lead to superior heat dissipation compared to that with Northerly winds.

In contrast, the SAT modules would experience boundary layer development over the front and rear surfaces during Southerly winds, as these would effectively act as flat plates relative to the wind direction. Based on forced convection heat transfer theory, heat dissipation from the SAT module surfaces would therefore be lower compared to the FT modules. A comparison of the left- and right-hand side of Figure 15 confirms enchanced heat dissipation factors for the FT modules during Southerly winds, which contributes to the slightly higher overall $U'_1$ value for this configuration.

Figure 15 also shows that during low-speed winds (0-1 m/s) from the West, the SAT modules experience greater heat dissipation than the FT setup. Our hypothesis is that enhanced heat dissipation is observed on the less inclined upper surface of the SAT modules due to more pronounced natural convection effects, as opposed to the heat transfer occurring across the comparatively stable boundary layer on the more inclined top surface of the FT module. As these conditions account for a significant portion of the overall data set, it results in the boosted $U_0'$ value for the SAT configuration compared to the FT setup.

Ultimately, the right-hand plot of Figure 14 indicates higher overall heat dissipation from the SAT modules over the test period.

### 3.4 Interactions between irradiance, wind speed and temperature difference

Figure 16 and Figure 17 display the relationships between heat dissipation, POA irradiance, wind speed and temperature difference for FT and SAT modules.

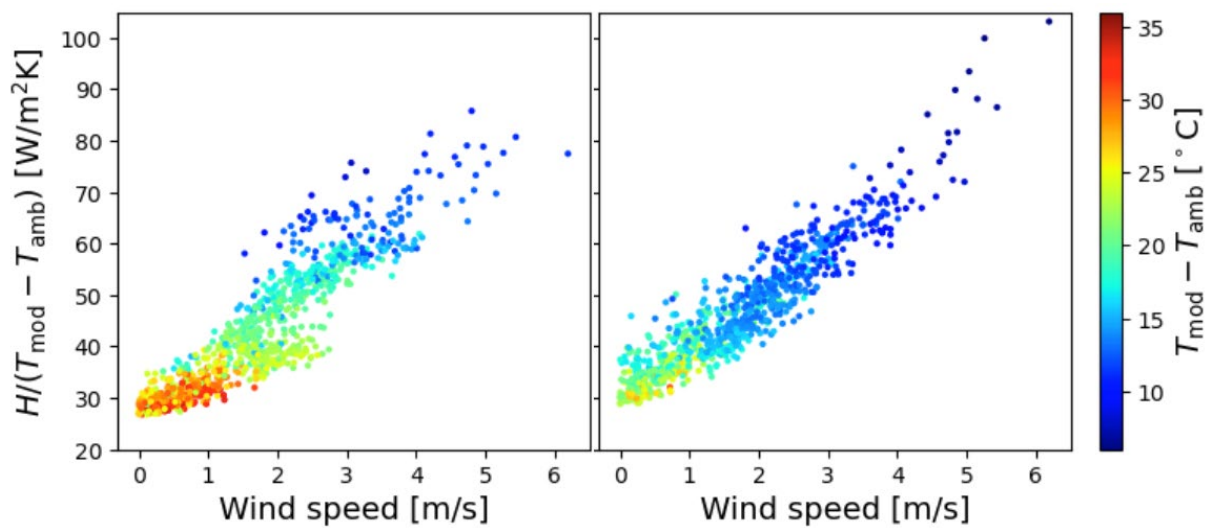

**Figure 16: Colour bar of temperature difference over the plot of H/(T$_{mod}$-T$_{amb}$) versus wind speed for the FT (left) and SAT (right) configurations**

Figure 16 clearly indicates the role that increasing wind speed has in reducing the temperature difference (effectively module temperature). Figure 17, in turn, shows a clear relationship between the temperature difference and POA irradiance. At constant wind speed, higher POA irradiance correlates to higher temperature difference. For constant POA irradiance, increasing wind speed causes a decreasing temperature difference. The temperature difference correlates well with different combinations of POA irradiance and wind speed.

As discussed in the previous section, these figures confirm the following interesting findings when comparing the FT and SAT configurations:

- Higher overall module operating temperatures for the FT modules, therefore enhanced heat dissipation from SAT modules. This is especially clear under no / low wind conditions.
- A greater level of data scatter for the FT modules, due to a larger range of irradiance incident on the North-facing surface as well as a sensitivity to wind direction.

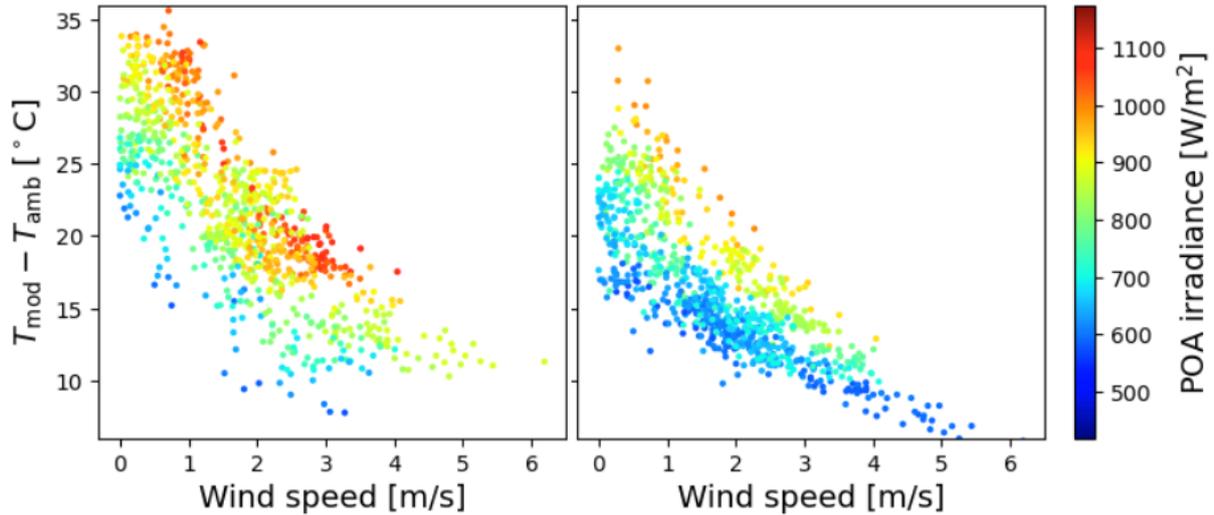

Figure 17: Colour bar of POA irradiance over the plot of H/($T_{mod}$-$T_{amb}$) versus wind speed for the FT (left) and SAT (right) configurations

## 3.5 Filtering effects

In this section, we address the impact of the chosen data filtering approach on the predicted heat dissipation. While this study utilized the filtering approach of Faiman [7] (as motivated in Section 2.4), others ([20, 22]) have employed the specifications of the IEC 61853-2 standard [21]. The standard specifically rejects data according to the following filters:

- Low irradiance filter: Irradiance values below 400 W/m$^2$.
- Irradiance fluctuations filter: Irradiance values for a 10-minute interval when the irradiance varies by more than 10% from the irradiance value during the preceding 10-minute interval.
- Wind fluctuations and gust filter: Wind speed values in a 10-minute interval which includes a deviation of the instantaneous wind speed below 0.25 m/s or gusts larger than +200% from the 10-min average.

To evaluate the influence of filtering approach, the complete FT and SAT 5-minute averaged data sets between 30/03/2023 and 07/08/2023 were filtered according to the above method. For the SAT configuration, periods of tracker malfunction were excluded from the pre-filtered data set. Figure 18 and Figure 19 present graphical comparisons of the remaining data after employing the filtering approach of [7] (a) and [21] (b). Of the available 34142 data points, for the FT configuration, 5166 points pass the IEC filters compared to 3378 for the Faiman filtering method. For the SAT configuration, of the available 24114 data points, 3072 points pass the IEC filters compared to 1746 points using the Faiman approach.

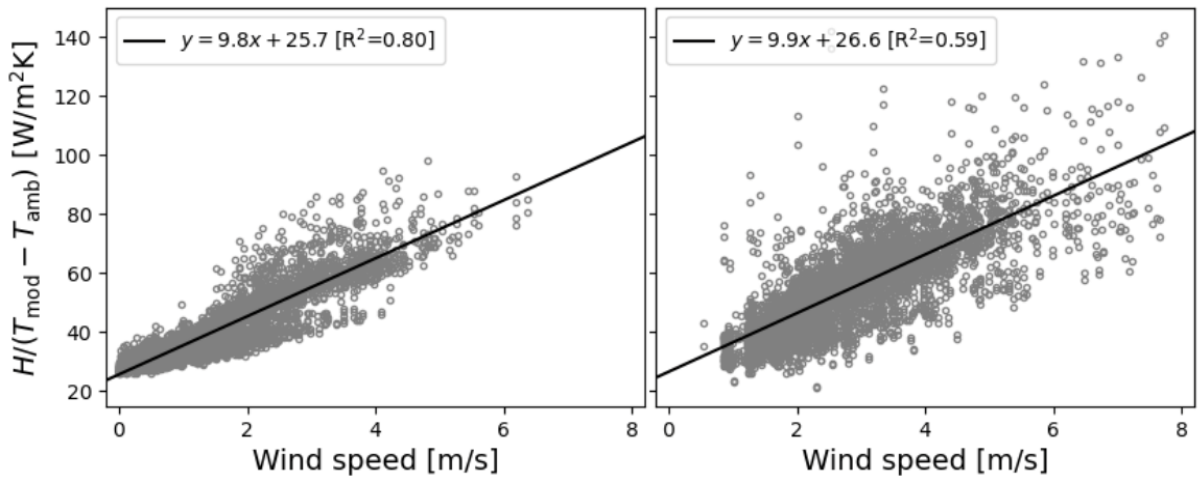

**Figure 18: Plot of H/($T_{mod} - T_{amb}$) versus wind speed for the FT configuration, and the associated linear fit for (a) all clear sky 5-minute data points (b) anytime data with IEC [21] filters**

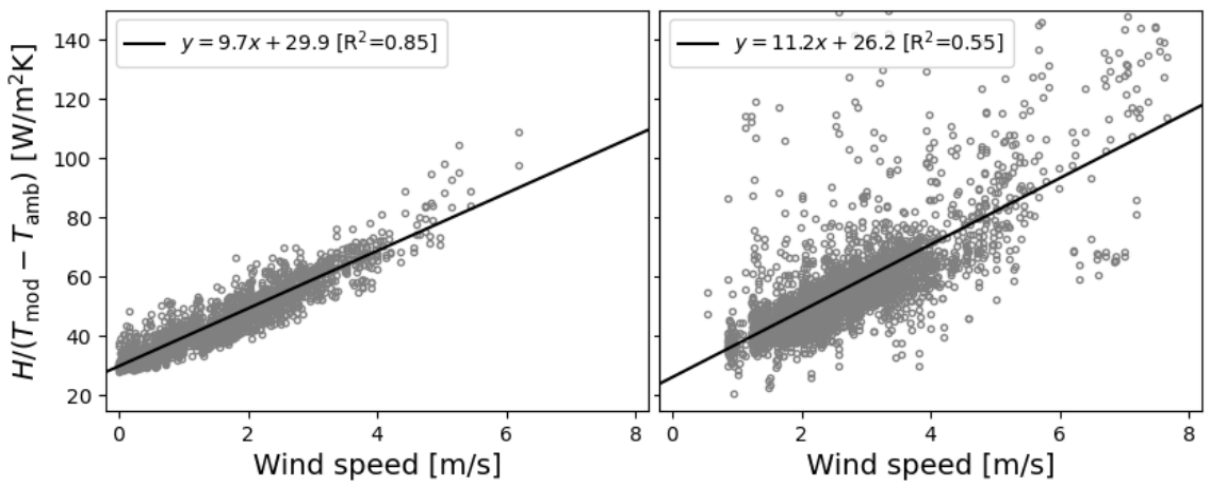

**Figure 19: Plot of H/($T_{mod} - T_{amb}$) versus wind speed for the SAT configuration, and the associated linear fit for (a) all clear sky 5-minute data points (b) anytime data with IEC [21] filters**

The IEC filtering method exhibits increased scatter because it encompasses data from diverse conditions throughout the testing period. The level of scatter amplification for the SAT configuration is more severe. While these modules may experience irradiance levels above the minimum limit, small temperature differences between the module and ambient are often encountered early and late in the day. This makes the heat dissipation factor values particularly sensitive and prone to scatter away from the linear trend. In contrast, the Faiman method isolates the most stable conditions for data collection. While the IEC method selectively filters out only the most adverse conditions, it is not stringent enough to minimize variability to the extent observed in the clear sky plot.

Table 3 shows a comparison of the heat dissipation factors derived from the data of Figure 18 and Figure 19. Despite the differences in scatter for the FT modules, the overall results are very similar. This is not quite the case for the SAT modules, where differences in overall heat dissipation factors are more significant.

**Table 3: Comparative heat dissipation factors for FT and SAT configurations, showing effect of data filtering approach**

| Filters | $U'_0$ (W/m²K) | $U'_1$ (Ws/m³K) | $R^2$ |
|---|---|---|---|
| FT clear sky (10:00-14:00) | 25.7 | 9.8 | 0.80 |
| FT IEC 61853-2 | 26.6 | 9.9 | 0.59 |
| SAT clear sky (10:00-14:00) | 29.9 | 9.7 | 0.85 |
| SAT IEC 61853-2 | 26.2 | 11.2 | 0.55 |

Table 4 displays the heat dissipation factors reported by various sources for FT open-rack PV module configurations. The factors obtained in this study, through either filtering approach, align well with those found in the literature. Except for [22], the $U'_0$ values exhibit remarkable similarity across sources, whereas a range of $U'_1$ values is observed. The heat loss factors employed by PVsyst for the simulation of open-rack installations are also included for comparison.

**Table 4: Heat dissipation factors and temperature prediction comparison across literature sources for FT configuration**

| Configuration | $U'_0$ (W/m²K) | $U'_1$ (Ws/m³K) | RMSE (°C) |
|---|---|---|---|
| Clear sky (10:00-14:00) | 25.7 | 9.8 | 2.39 |
| IEC 61853-2 | 26.6 | 9.9 | 2.38 |
| Owen et al. [24] 23° tilt | 26.5 | 5.2 | 2.82 |
| Faiman [7] | 25.0 | 6.84 | 2.65 |
| Koehl et al. [20] Negev year 2 | 26.9 | 6.1 | 2.57 |
| Barykina and Hammer [22] | 34.5 | 4.44 | 2.52 |
| PVsyst [14]* | 41.4 | 0 | 3.07 |

*Assuming $\eta_o = 0.8$ and $\eta_e = 0.1$, as per [7]

To determine the sensitivity of module temperature prediction to specific heat dissipation factors, the following evaluation is performed. The evaluation is limited to the FT configuration only, due to the availability of published heat dissipation factors from other researchers. Based on the current study's unfiltered data set (representing day and nighttime), the heat dissipation factors for each of the literature sources are used in turn to predict module temperature. The actual measured module temperatures are then compared to the predictions and a RMSE value is calculated for each model. It is expected that the heat dissipation factors defined by this work provide the best RMSE, since they are derived from the data set of measured temperature used in the prediction. Yet, results show that the heat dissipation factors from literature provide very similar results in terms of RMSE, even with the range of $U_0'$ and $U_1'$ values noted. It is interesting to note that the predictions by PVsyst show the greatest error.

Three noteworthy conclusions can be drawn from these results. First, although the IEC filtering method had a more scattered plot and lower $R^2$ value of fit compared to Faiman's filtering approach, there was no inaccuracy as a result. Therefore, using the simpler Faiman method of filtering instead of the IEC filters shows no loss in accuracy when predicting the module temperature.

Second, except for PVsyst, the heat dissipation factors from all relevant sources predicted module temperature for the FT open-rack configuration within 0.5°C of each other. This seems to suggest that the exact values of the heat dissipation factors are not crucial in the accuracy of predicting module temperature for FT open-rack modules. We believe that this insensitivity stems from the high level of exposure of both surfaces of open-rack installations to the environment. It would be our expectation that different BAPV systems, for example, would show greater sensitivity to the specific configuration and local conditions of the particular setup. It should however be noted that our above analysis considered heat dissipation factors within certain limits, as well as a single data set, and it is consequently not clear whether this finding will always be valid for FT open-rack systems.

Thirdly, results suggest that it is preferable to consider both heat dissipation factors for prediction of module temperatures instead of an inflated $U_0'$ value only, even if accurate information is not available.

**3.6 Daily variation effects**

Among the 44 days with clear skies, Figure 20 highlights data from three specific days for the FT configuration, each distinguished by colour. The figure scrutinizes the behaviour and impact of these days on the data set. Notably, two outlier days (depicted in red and blue) and one day which aligns with the linear trend (depicted in green) are observed.

For the day marked by green indicators, the rising trend in heat dissipation factors corresponds to a gradual increase in wind speed spanning a considerable range throughout the day. During this period, there is minimal fluctuation in instantaneous wind speed, resulting in limited gusts and consequently, little variation in heat dissipation factors around the trendline.

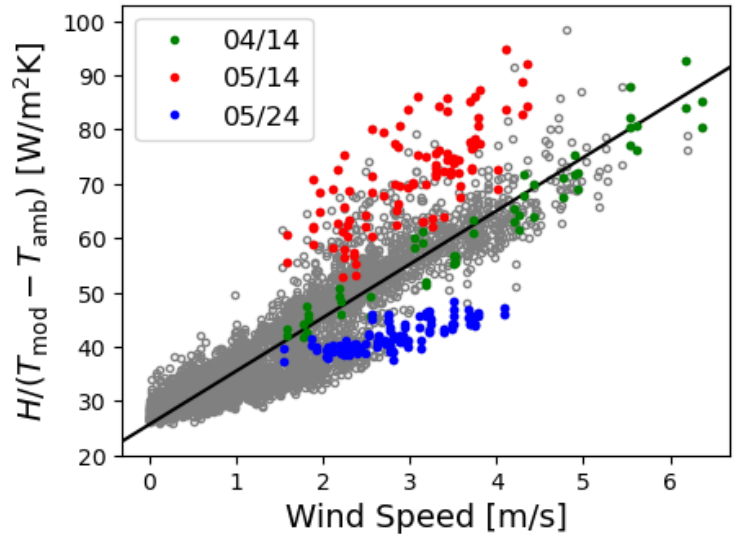

**Figure 20: Daily variation in heat dissipation from FT modules for three selected days**

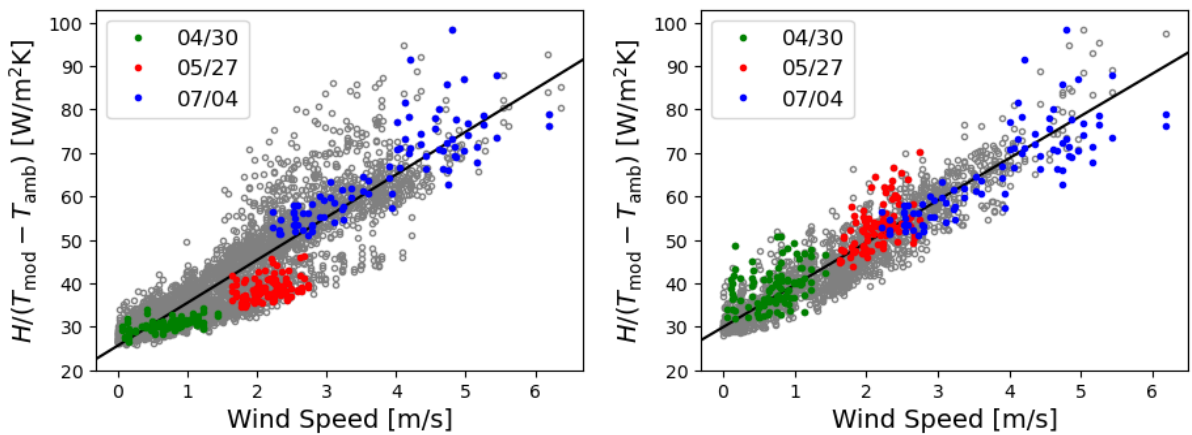

**Figure 21: Comparison of daily variation in heat dissipation from FT (left) and SAT (right) modules over the same three days**

The days represented by red and blue data exhibit similar magnitudes of POA irradiance, wind speed, and ambient temperature. However, there is a notable difference in wind speed fluctuation between the red and blue days, contributing to distinct scatter patterns on the graph. Despite comparable atmospheric conditions, module temperatures are higher for the blue data. Consequently, the ratio of $H$ to the temperature difference $(T_{mod} - T_{amb})$ is lower for the blue data due to the higher denominator, causing the data points to lie below the trendline. The variance in module temperature between these two days is attributed to differences in wind direction.

Figure 21 presents a comparison of daily variations on heat dissipation for the FT and SAT modules, over the same three days. The selected days are different to those in Figure 20 due to tracker limitations for the SAT configuration on the indicated days, which did not allow for comparison. The data presented in Figure 21 clearly indicates that, during days with low wind conditions, the FT modules exhibit a deviation below the established trendline. Conversely, under identical ambient conditions, the SAT modules consistently conform to the linear trendline. In instances of elevated wind speeds, the data for both configurations tend to be in closer proximity to the trendline. These observations again highlight the variability in heat dissipation from FT modules, the enhanced heat dissipation capability of the SAT modules during low wind, and the relatively similar heat losses during higher wind conditions.

## 4. Implication of results

The implications of the configuration-specific heat dissipation factors which were obtained in this study are investigated through simulation. The annual electricity output of systems with each mounting configuration is determined using PVsyst version 7.4. Equations 2 and 3 are used, together with the assumptions noted below, to convert the accented heat dissipation factors to the $U_0$ and $U_1$ input values that PVsyst requires. For each configuration, the following scenarios were simulated:

1. Modelled configuration using default heat dissipation factors defined by PVsyst.
2. Modelled configuration using the configuration-specific heat dissipation factors obtained in this study.

PVsyst has no values specifically for the SAT configuration, and only presents values for open-rack applications. As such, the simulations below employed the PVsyst default open-rack heat dissipation factors as inputs for both FT and SAT scenarios. The mounting / tracking input parameters are detailed in Table 5. The overall system parameters chosen for the FT/SAT simulations are given in Table 6.

**Table 5: FT and SAT mounting / tracking input parameters**

|                       | FT   | SAT   |
|-----------------------|------|-------|
| Tilt angle            | 31°  | 0°    |
| Module azimuth angle  | 0°   | 0°    |
| PV tracker min. angle | -    | -50°  |
| PV tracker max. angle | -    | 50°   |

**Table 6: FT / SAT system simulation input parameters**

| Parameter | Value |
|---|---|
| PV module | CS3W-420P 1500 V |
| Optical efficiency ($\eta_o$) | 0.8 |
| Electrical efficiency ($\eta_e$) | 0.1 |
| Inverter | Fronius Symo 6.0 kW |
| Geographical site | Stellenbosch, South Africa |
| Nb of module in series | 12 |
| Nb of strings | 1 |

The optical efficiency of the PV module is in the range of 0.8–0.9 [7]. The PV module has an electrical efficiency $\eta_e$ of 0.19 under Standard Test Conditions (STC). Faiman [7] noted an efficiency around 0.1 for normal operating temperatures. Based on these values, the difference between optical efficiency and electrical efficiency, $\eta_o - \eta_e$ used in the simulations was 0.7. The ~5 kW PV system sizing was chosen for its modelling simplicity. The system allows the PV array to be connected to one string and connected to an inverter where there is no limit to the annual energy generated.

The heat dissipation factors for the FT and SAT simulation scenarios and the annual electricity output for each are given in Table 7.

**Table 7: Heat dissipation factor inputs and annual output comparison**

| Parameter | FT | | SAT | |
|---|---|---|---|---|
| | 1 | 2 | 1 | 2 |
| $U_0$ (W/m²K) | 29 | 17.99 | 29 | 20.93 |
| $U_1$ (Ws/m³K) | 0 | 6.86 | 0 | 6.79 |
| Annual electricity output (kWh) | 9899 | 10190 | 10634 | 10989 |

Simulation results show increases in predicted annual output of 2.9% and 3.3% for FT and SAT open-rack systems respectively, when implementing configuration-specific heat dissipation factors compared to PVsyst defaults. These results are significant, especially when viewed in the light of increasing competition between bids for construction of modern PV power plants and the required accuracy of techno-economic calculations. These findings emphasize the advantages of more realistic and configuration-specific heat dissipation factors.

## 5. Conclusion

This study conducted long-term experiments for FT and SAT open-rack PV modules at a site in South Africa. The heat dissipation model and data filtering method of Faiman was employed, which entails the consideration of clear sky day data only between 10:00 and 14:00.

The results of the FT experiments compare favorably to results from literature, producing $U'_0 = 25.7$ W/m$^2$K and $U'_1 = 9.8$ Ws/m$^3$K from the linear data regression analysis for the heat dissipation factors. Experiments on SAT modules extract heat dissipation factors of $U'_0 = 29.9$ W/m$^2$K and $U'_1 = 9.7$ Ws/m$^3$K, which represent a novel contribution in the field. These values indicate that enhanced heat dissipation is experienced at no / low wind conditions for SAT modules compared to FT modules.

An exploration into selected days and averaged data over the test period reveals a clear relationship between module temperature, POA irradiance and wind speed. Evaluations plainly confirm lower module temperatures due to boosted heat dissipation from SAT modules in relation to FT modules.

Interestingly, wind direction plays a secondary but not insignificant role in the heat dissipation of FT modules, while heat dissipation from SAT modules seems to be insensitive to wind direction. Greater variability in heat dissipation is observed from FT modules compared to SAT modules, due to the larger range of incident irradiance onto the FT surfaces and the sensitivity to wind direction.

An investigation examining the impact of data filtering methods on heat dissipation characteristics evaluated the methodologies proposed by Faiman and the IEC 61853-2 standard. It concludes that the heat dissipation factors for SAT modules are somewhat more sensitive to the specific filtering method employed compared to FT modules.

A further analysis on FT systems sequentially employs heat dissipation factors from various sources for predicting module temperature based on the current study's full data set. RMSE values for each model fall within 0.5°C of each other, indicating that precise values of heat dissipation factors are likely not critical for reliable forecasts of module temperature for FT open-rack systems. It is expected that the same will not be true for PV systems which are less exposed to the ambient environment, such as BAPV configurations.

The study finally conducts annual power output simulations using the commercial software PVsyst. Simulations evaluate FT and SAT configurations, comparing performance when implementing PVsyst's default heat dissipation factors versus the experimentally determined heat dissipation factors from the study. Results show a 2.9% and 3.3% enhancement in annual power output for the FT and SAT configurations respectively when utilizing the experimental heat dissipation factors. These findings carry significance, particularly in the context of growing competition in tendering for the construction of contemporary PV power plants and the heightened importance of precise techno-economic calculations. The results underscore the benefits of utilizing heat dissipation factors that are realistic and tailored to the specific configuration.

**Nomenclature**

**Variables**

| | |
|---|---|
| $H$ | Plane-of-array irradiance (W/m$^2$) |
| $R^2$ | Coefficient of determination |
| $T$ | Temperature (°C) |
| $U$ | Heat dissipation factor (W/m$^2$K or Ws/m$^3$K) |
| $V$ | Speed (m/s) |

**Subscripts**

| | |
|---|---|
| $0$ | Constant |
| $1$ | Wind-dependant |
| $amb$ | Ambient |
| $e$ | Electrical |
| $mod$ | PV module |
| $o$ | Optical |
| $w$ | Wind |

**Superscripts**

| | |
|---|---|
| ' | Modified |

**Greek symbols**

| | |
|---|---|
| $\eta$ | Efficiency (%) |

**Abbreviations**

| | |
|---|---|
| BAPV | Building-attached photovoltaic |
| BIPV | Building-integrated photovoltaic |
| FT | Fixed-tilt |
| HAGL | Height above ground level |
| HAMSL | Height above mean sea level |
| IEC | International Electrotechnical Commission |
| MPPT | Maximum power point tracking |
| PC | Personal computer |
| POA | Plane-of-array |
| PV | Photovoltaic |
| PWM | Pulse width modulation |
| RMSE | Root mean square error |
| SAT | Single-axis tracked |
| SUNREC | Stellenbosch University Renewable Energy Center |
| UPS | Uninterruptible power supply |
| USB | Universal Serial Bus |

**Data availability statement**

The data supporting the findings of this study are available upon reasonable request from the corresponding author.

**Funding statement**

No specific funding was utilized for the completion of this research.

**Conflict of interest disclosure**

The authors confirm that no conflicts of interest exist.

**Acknowledgements**

The authors would like to thank the Thermofluids division of the Mechanical and Mechatronic Engineering Department at Stellenbosch University for the sponsorship of test equipment used in this study.